\documentclass[aps,pre,twocolumn,groupedaddress,amsmath,epsfig,amssymb]{revtex4}
\usepackage{makecell}
\usepackage{bbm}
\usepackage{bm}
\usepackage{mathrsfs}
\usepackage{amsfonts}
\usepackage{color}
 \usepackage{hyperref} 
 \hypersetup{
 colorlinks=true,
 linkcolor=blue,
 urlcolor=blue,
 }
\usepackage{mathtools, nccmath}
\usepackage{graphicx}
\newcommand{\dbar}{d\hspace*{-0.08em}\bar{}\hspace*{0.1em}}

\newcommand{\wu }[1] {{\color{magenta} #1 }}

\begin{document}
\newcommand{\Rv}{{\vec {R}}}
\newcommand{\rv}{{\vec r}}
\newcommand{\tv}{{\vec t}}
\newcommand{\av}{\boldsymbol a}
\newcommand{\fv}{{\boldsymbol f}}
\newcommand{\hv}{{\boldsymbol h}}
\newcommand{\xv}{{\boldsymbol x}}
\newcommand{\xp}{\vec{x}_{\perp}}
\newcommand{\pp}{\vec{p}_{\perp}}
\newcommand{\zv}{{\boldsymbol z}}
\newcommand{\uv}{{\vec u}}
\newcommand{\Av}{{\boldsymbol A}}
\newcommand{\Xv}{{\boldsymbol X}}
\newcommand{\Yv}{{\boldsymbol Y}}
\newcommand{\Pv}{{\boldsymbol P}}
\newcommand{\Qv}{{\boldsymbol Q}}
\newcommand{\ur}{\vec{{\EuFrak u}}}
\newcommand{\cv}{{\vec c}}
\newcommand{\qv}{{\boldsymbol q}}
\newcommand{\pv}{{\boldsymbol p}}
\newcommand{\yv}{{\boldsymbol y}}
\newcommand{\vv}{{\boldsymbol v}}
\newcommand{\kv}{{\vec k}}
\newcommand{\phiv}{{\boldsymbol \phi}}
\newcommand{\etav}{{\boldsymbol \eta}}
\newcommand{\Tr}{{\rm Tr}}
\newcommand{\px}{{\partial_x}}
\newcommand{\py}{{\partial_y}}
\newcommand{\ppi}{{\partial_i}}
\newcommand{\ppj}{{\partial_j}}
\newcommand{\ch}{{\hat{c}}}
\newcommand{\eh}{{\hat{e}}}
\newcommand{\xh}{{\hat{x}}}
\newcommand{\yh}{{\hat{y}}}
\newcommand{\zh}{{\hat{z}}}
\newcommand{\vh}{{\hat{v}}}
\newcommand{\qh}{{\hat{q}}}
\newcommand{\kh}{{\hat{k}}}
\newcommand{\llm}{{\boldsymbol \lambda}}
\newcommand{\Am}{{\boldsymbol{A}}}
\newcommand{\Qm}{{\boldsymbol{Q}}}
\newcommand{\Rm}{{\boldsymbol R}}
\newcommand{\Lm}{{\boldsymbol L}}
\newcommand{\Km}{{\boldsymbol K}}
\newcommand{\Jm}{{\boldsymbol J}}
\newcommand{\Tm}{{\boldsymbol T}}
\newcommand{\Bm}{{\boldsymbol B}}
\newcommand{\Dm}{{\boldsymbol D}}
\newcommand{\Cm}{{\boldsymbol C}}
\newcommand{\Em}{{\boldsymbol E}}
\newcommand{\Mm}{{\mathcal M}}
\newcommand{\Wm}{{\boldsymbol W}}
\newcommand{\Fm}{{\boldsymbol F}}
\newcommand{\Gm}{{\boldsymbol G}}
\newcommand{\Imm}{{\boldsymbol I}}
\newcommand{\sm}{{\boldsymbol s}}
\newcommand{\gammam}{{\boldsymbol \gamma} }
\newcommand{\chim}{{\boldsymbol \chi}}
\newcommand{\be}{\begin{equation}}
\newcommand{\ee}{\end{equation}}
\newcommand{\ba}{\begin{eqnarray}}
\newcommand{\ea}{\end{eqnarray}}
\newcommand{\RNum}[1]{\uppercase\expandafter{\romannumeral #1\relax}}
\newcommand{\ddelta}{\boldsymbol{\delta}}
\newcommand{\pdf}{p}
\newcommand{\LFP}{\mathcal{L}_{\rm FP}}

\title{{Stochastic Thermodynamics of Brownian motion in Temperature Gradient}}
\author{Mingnan Ding$^{1}$}
\email{dmnphy@sjtu.edu.cn}
\author{Jun Wu$^{1}$}
\author{Xiangjun Xing$^{1,2,3}$}
\email{xxing@sjtu.edu.cn}
\affiliation{$^1$Wilczek Quantum Center, School of Physics and Astronomy, Shanghai Jiao Tong University, Shanghai, 200240 China\\
$^2$T.D. Lee Institute, Shanghai Jiao Tong University, Shanghai, 200240 China\\
$^3$Shanghai Research Center for Quantum Sciences, Shanghai 201315, China}
\date{\today} 

\begin{abstract}  
{We study stochastic thermodynamics of a Brownian particle which is subjected to  a temperature gradient and is confined by an external potential.  We first formulate an over-damped Ito-Langevin theory in terms of local temperature, friction coefficient, and steady state distribution, all of which are experimentally measurable.   We then study the associated stochastic thermodynamics theory.  We analyze the excess entropy production (EP) both at trajectory level and at ensemble level, and derive the Clausius inequality as well as the transient fluctuation theorem (FT).  We also use molecular dynamics to simulate a Brownian particle inside a Lennard-Jones fluid and verify the FT.  Remarkably we find that the FT remains valid even in  the under-damped regime.  We explain the possible mechanism underlying this surprising result. }

\end{abstract}

\maketitle
\section{ Introduction} 
After several decades of hard work, two types of Fluctuation theorems (FTs)~\cite{Gallavotti-1995,Searles1999,Crooks1999,Seifert2012,Jarzynski2011,Ciliberto-2017}  have been firmly established.  Firstly the {steady state FT} (SSFT) characterizes the large deviation behaviors~\cite{Searles1999,Zon2003,Lebowitz-1999,Gallavotti-1995} of entropy production (EP) rate in non-equilibrium steady states.  Secondly, various {transient FTs}~\cite{Crooks1999,Jarzynski2011,Seifert2012} characterize the fluctuations of dissipated work in non-equilibrium processes starting from equilibrium states.  These results constitute the backbone of an emerging field called {\em stochastic thermodynamics}~\cite{Seifert2012,Peliti-Pigolotti-2021,Broeck-intro-2013}, which provides a unified view of non-equilibrium fluctuations.   

Many physical systems and most biological systems are however embedded in dissipative backgrounds and  undergo non-stationary processes.  They therefore do not fall into either of above mentioned two categories.  Remarkably, a series of seminal theoretical works~\cite{Hatano2001,Speck-Seifert-2005,Jarzynski-2006,Esposito2010,Esposito2010-2,Broeck2010,Ge-Qian-2010} indicate that in Markov processes with even variables, the total EP may be broken into two parts, each satisfying a separate FT.  For Langevin systems, it may be understood that these two parts of EP are due  to  respectively the dissipative work of the conservative and non-conservative components of forces~\cite{nc-2022}.  This decomposition of EP is also intimately related to the Glansdorff-Prigogine stability theory of non-equilibrium steady states~\cite{Glansdorff-Prigogine-1971,Qian-Hong-2001,Qian-Hong-2002,nc-2022}, as well as the {\em steady-state thermodynamics} of Sasa and Tasaki~\cite{Sasa-Tasaki-2006-SST,nc-2022}. Following these earlier literatures, it is therefore natural to call them respectively the excess EP and the house-keeping EP~\footnote{They are called respectively {\em the non-adiabatic EP} and {\em adiabatic EP} by Esposito and Van den Broeck~\cite{Esposito2010,Esposito2010-2,Broeck2010}.}. 
We note that except for a few partial results~\cite{Trepagnier2004,Thalheim-2020-Soret-EQ}, there has been no systematic verification of these theories, either experimentally or numerically.  

Brownian motion in a temperature gradient constitutes one of the simplest  systems embedded in dissipative backgrounds.  In the absence of external force, a particle immersed in a temperature exhibits directed motion, an effect known as {\em Ludwig-Soret effect}~\cite{Ludwig-1859,Soret-1879,Groot-1945}, or {\em thermophoresis}~\cite{Piazza-review-2008,Piazza-soft-matter-2008,Groot-non-EQ-TM}, which has many important applications and has been studied in various situations~\cite{Willemsen-2014,Willemsen-2011,Thamdrup-2010,Baaske-2010,wen_temperature_2020,becton_thermal_2014,Dehbi2009,Pakravan2011,Helden2015,Braibanti2008,rurali_thermally_2010,schoen_phonon_2007,duhr_thermophoretic_2006,duhr_why_2006,Kroy-2016-hot-Brownian,Falasco2016-2}.  Regardless of significant recent progresses~\cite{Wurger2016,Wuger2016-2,Burelbach}, the statistical mechanism of thermophoresis is not yet fully understood~\cite{Piazza-review-2008,Falasco2016-1}.  The stochastic thermodynamics of Brownian motion in temperature gradient has also attracted some recent interests~\cite{Celani2012,Marino2016, Sancho2015,Polettini2013-2}.  Due to the difficulties associated with the multiplicative nature of noises, however, a systematic and consistent theory however has yet to be developed.  In particular there has been no derivation or verification of fluctuation theorems for Brownian motion in temperature gradient. 


In the present work, we shall develop a systematic theory of stochastic thermodynamics for this system, and make contact with thermodynamics of thermophoresis~\cite{Groot-non-EQ-TM,Piazza-review-2008}.    In Sec.~\ref{sec:Langevin-eq},  we shall model the over-damped Brownian dynamics using covariant Langevin equation, whose general theory was developed in Ref.~\cite{covariant-Langevin-2020}.  Unlike the previous studies~\cite{Celani2012,Marino2016, Sancho2015,Polettini2013-2}, our Langevin equation is fully characterized by steady-state distribution and the position dependent temperature/friction coefficient, all of which are experimentally measurable. Regardless of the non-equilibrium nature of the background fluid, however, the Langevin equation obeys the conditions of detailed balance.  In Sec.~\ref{sec:st}, we proceed to construct a stochastic thermodynamic theory using the over-damped Langevin equation.  Because the background does not have a fixed temperature, however, the conventional formulation of stochastic thermodynamics in terms of heat and work is not applicable.  We therefore forego the concepts of work/heat and deal with EP directly.  Also the background fluid is in a non-equilibrium steady state which constantly dissipates energy.  This dissipation is completely invisible in our theory.  Hence the EP captured by our theory is not the full EP, but only {\em the excess EP}.  We study this excess EP both at the trajectory level and at the ensemble level, and prove rigorously the Clausius inequality as well as the transient FT obeyed by this quantity.  In the limit of vanishing temperature gradient, our theory reduces to the usual theory of stochastic thermodynamics.   Our theory can be understood as a simple showcase of the general theory which we developed in Ref.~\cite{nc-2022}.  In Sec.~\ref{sec:simulation}, we verify our theory by simulating a Brownian particle immersed in a Lennard-Jones fluid subjected to a temperature gradient.  We compute the position-dependent temperature and friction coefficient, as well as the generalized potential using the simulation data.  Using these results, we explicitly verify the validity of the FT for the excess EP.  Remarkably, the FT remains valid not only in the over-damped regime, but also deep in the under-damped regime.  We briefly explain this surprising result.  Finally in Sec.~\ref{sec:conclusion} we summarize our result and project future directions. 


\section{ Langevin dynamics} 
\label{sec:Langevin-eq}

As discussed in the Introduction, we start directly with the over-damped limit.  Let $\xv$ be the position vector of the Brownian particle, $T(\xv), \gamma(\xv)$  respectively the position-dependent temperature and friction coefficient.  We assume that the Brownian particle is coupled to a confining potential, so that its steady state probability density function (pdf) $ \pdf_{\rm ss}(\xv)$ is normalizable.  Defining the {\em generalized potential} $U(\xv)$ via 
 \ba
 \pdf_{\rm ss}(\xv) \equiv e^{-U(\xv)},
 \label{p_ss-U-1}
 \ea
the over-damped Langevin equation is given by
\ba
 dx^i =- \frac{T(\xv)}{\gamma(\xv)} \partial_i U(\xv)dt
  + \partial_i \frac{T(\xv)}{\gamma(\xv)}  dt
+ \sqrt{ \frac{2 T(\xv)}{\gamma(\xv)}}dW^i,
\quad \label{od-Ito-Langevin}
\ea
where $x^i$ is i-th Cartesian component of  $\xv$, $\partial_i = \partial /\partial x^i$, whilst $dW^i$ are the standard Wiener noises satisfying
 \begin{subequations}
\ba
\langle dW^i \rangle &=&0, \\
\langle dW^i dW^j \rangle &=&\delta^{ij} \, dt. 
\ea
\label{Wiener-noises}
\end{subequations}
\vspace{-3mm}

The ratio $T(\xv)/\gamma(\xv)$ can be understood as the position-dependent diffusion constant. The product $ \sqrt{ {2 T}/{\gamma }} \, dW^i$ in the r.h.s. of Eq.~(\ref{od-Ito-Langevin}) is defined in Ito's sense.    Letting $p = p(\xv,t)$ be the pdf of $\xv$, using the standard methods of stochastic analysis~\cite{Gardiner-book}, one can show that the Langevin equation (\ref{od-Ito-Langevin}) is mathematically equivalent to the following {\em Fokker-Planck equation} (FPE):
\ba
\partial_t \, \pdf = \partial_i \frac{T(\xv)}{\gamma(\xv)} 
\left( \partial_i+ \left(\partial_i U \right) \right) \pdf. 
\label{FPE}
\ea
 Note that repeated indices are summed over.   In this work we do not distinguish superscripts from subscripts.  This does not lead to any ambiguity, because we will not carry out any nonlinear coordinate transformation.  It is easily seen that Eq.~(\ref{p_ss-U-1}) is indeed the steady state solution of Eq.~(\ref{FPE}).  Note that the FPE (\ref{FPE}) can also be written as
\ba
\partial_t \, \pdf = - \partial_i j_i, 
\label{FPE-j}
\ea
where $j_i = j_i(\xv, t) $ is the probability current:
\ba
j_i(\xv, t) = - \frac{T(\xv)}{\gamma(\xv)} (\partial_i + (\partial_i U) ) p(\xv,t),
\label{j_2}
\ea
which vanishes identically in the steady state Eq.~(\ref{p_ss-U-1}).

The Langevin equation  (\ref{od-Ito-Langevin}) and the Fokker-Planck equation (\ref{FPE}) are in the covariant form as studied in Ref.~\cite{covariant-Langevin-2020}, with a diagonal but position dependent kinetic matrix $L^{ij}(\xv) = (T(\xv)/\gamma(\xv)) \delta^{ij}$.  The term $\partial_i ({T(\xv)}/{\gamma(\xv)}) dt $ in the r.h.s. of Eq.~(\ref{od-Ito-Langevin}) is called the {\em spurious drift}~\cite{covariant-Langevin-2020}, which is necessary to guarantee that  Eqs.~(\ref{od-Ito-Langevin}) and (\ref{FPE}) are mathematically equivalent.

Since the position variable $\xv$  is even under time reversal, Eq.~(\ref{od-Ito-Langevin}) also satisfies {\em conditions of detailed balance}, which are given in Eqs.~(3.16) of Ref.~\cite{covariant-Langevin-2020}, or Eqs. (2.31) of Ref.~\cite{covariant-ST-2021}.  This guarantees that two-time pdf satisfies the following symmetry: 
$$p(\xv_2, t_2; \xv_1, t_1) =p(\xv_1, t_1;\xv_2, t_2), $$ 
where $\xv_1, \xv_2$ are two distinct position vectors.




The generalized potential $U(\xv)$, whose concrete form is irrelevant at this stage, depends both on the external confining potential and on the temperature gradient.  In the absence of external confining potential, the Brownian particle exhibits thermophoresis, i.e., it moves with a constant average speed, parallel or anti-parallel to the temperature gradient.  This implies that $U(\xv)$ has a term approximately linear in temperature $T(\xv)$, which will be verified using simulation data in  Sec.~\ref{sec:property-brownian-particle-U}.  Additionally, as usual in the study of stochastic thermodynamics, the external confining potential is externally tuned by certain control parameters. These parameters are defined and discussed in detail in Sec.~\ref{sec:simulation}.  As a consequence, we should write the generalized potential as $U(\xv; \lambda)$, and the external confining potential as $V(\xv; \lambda)$, where $\lambda$ denote the control parameters.  To avoid  clutter, however, we  often hide the dependence of $V$ and $U$ on $\xv$ and $\lambda$. 


 
 \section{Stochastic thermodynamics}
 \label{sec:st}

In Ref.~\cite{covariant-ST-2021} we constructed a theory of stochastic thermodynamics for covariant nonlinear Langevin systems satisfying detailed balance~\cite{covariant-Langevin-2020}, under the assumption that the system is in contact with a heat bath with a fixed temperature.  The quantity that plays a crucial role in this theory is {\em Hamiltonian of mean force} (HMF) $ H(\xv; \lambda)$, which is related to the equilibrium pdf via:
\ba
p^{\rm eq}(\xv; \lambda) = e^{- U(\xv; \lambda)} = e^{\beta (F -  H(\xv; \lambda))},
\label{p-eq-x-lambda-H-F}
\ea
where $U(\xv; \lambda)$ is the generalized potential, and $F = F(\lambda)$ is the equilibrium free energy defined as
\ba
F(\lambda) = - T \log \int_\xv e^{-\beta  H(\xv; \lambda)}.
\label{F-T-int-x-exp}
\ea  
Note that we use $\int_{\boldsymbol x} \equiv \int d{\boldsymbol x}$ to denote volume integral over $\boldsymbol x$.  As one can see from Eq.~(\ref{p-eq-x-lambda-H-F}), both $ H(\xv; \lambda)$ and $F(\lambda)$ are defined only up to an additive constant, which may depend on $\lambda$.  This uncertainty arises because the microscopic Hamiltonian is defined only up to an additive constant.  We then defined~\cite{covariant-ST-2021} $ H(\xv; \lambda)$ as the fluctuating internal energy, and defined differential heat and work at the trajectory level as:
$ \dbar {\mathscr W} \equiv d_{\lambda}  H(\xv; \lambda),  
  \dbar {\mathscr Q} \equiv d_{\xv}  H(\xv; \lambda)$, where $d_\lambda, d_\xv$ are respectively the differentials due to variations of $\lambda$ and of $\xv$.   {Entropy} production was constructed using these quantities. The first law, the second law, as well as various FTs were then established.    

The approach {pursued in Ref.~\cite{covariant-ST-2021} is however not directly applicable} in the present case.   We note that, in the presence of temperature gradient, the r.h.s. of Eq.~(\ref{F-T-int-x-exp}) becomes $\xv$-dependent, whereas the l.h.s. is expected to be $\xv$-independent.  Hence there is no straightforward way to define free energy of a Brownian particle if temperature is position dependent.  The same difficulty also shows up when one tries to define HMF, or the differential heat and work.   



In view of this difficulty, we shall avoid defining any energy-like quantity, and deal only with entropy-like quantities.  This does not prevent us from understanding of the second law of thermodynamics or FTs, since the essence of these results is entropy rather than energy.  

As explained in the Introduction, the EPs that we shall study using the Langevin theory (\ref{od-Ito-Langevin}) is the {\em excess EP} which arises due to the insertion of the Brownian particle and variation of external control parameter.  The missing entropy product, which may be called the {\em housekeeping EP}, is extensive in the size of the fluid and is completely invisible in our theory.

\subsection{ Excess EP and excess Clausius inequality}

As in previous works~\cite{covariant-Langevin-2020,covariant-ST-2021}, we introduce the notations for partial differentials:
\ba
d_{\xv} U (\xv; \lambda) &\equiv& U(\xv + d \xv, \lambda ) 
- U(\xv; \lambda),
\label{d_xU-def}
\\
d_{\lambda} U (\xv; \lambda) 
&\equiv& U(\xv, \lambda + d \lambda ) - U(\xv; \lambda). 
\label{d_lambdaU-def}
\ea
The full differential of $U(\xv; \lambda)$ can be decomposed as
\ba
d U(\xv; \lambda) = d_{\xv} U (\xv; \lambda) 
+ d_{\lambda} U (\xv; \lambda). 
\label{dU-expand} 
\ea
It is important to note that $d_{\xv} U (\xv; \lambda) $ should be expanded in terms of $d\xv$ up to the second order:
\ba
d_{\xv} U (\xv)  = (\partial_i U) \, dx^i
 + \frac{1}{2} (\partial_i \partial_j U ) \, dx^i dx^j 
+ o(d\xv^2), 
\label{d_x-U-dx-1}
\ea
where all partial derivatives are evaluated at $\xv$.   It is well known that in Langevin dynamics with white noises $dW^i$ scales with $\sqrt{dt}$, whereas according to (\ref{od-Ito-Langevin}), we have $dx^i\sim dW^i \sim \sqrt{dt}$.   Hence expanding $d_{\xv} U (\xv) $ to the order $d\xv^2$ means that we are expanding it to the order $dt$, which is necessary 
in order to obtain a well-defined continuum limit.


We define the {\em excess} change of environmental entropy~\footnote{{As discuss in the Introduction, there is a housekeeping part of the change of the environmental entropy, which cannot be captured by our theory.  In an expanded theory which includes both the Brownian particle and the ambient fluid, both parts of entropy change show up explicitly.}  } at the trajectory level:
\ba
 d\mathcal S_{\rm env}^{\rm ex} \equiv - d_{\xv} U(\xv;\lambda),
\label{env-ep}
\ea
which depends both on the initial position $\xv$ and the incremental $d\xv$.  We may use Eq.~(\ref{d_x-U-dx-1}) to express $ d\mathcal S_{\rm env}^{\rm ex}$ in terms of $d\xv$, and use the Langevin equation (\ref{od-Ito-Langevin}) to express it in terms of Wiener noises $dW$.  Further averaging over the noises using Eqs.~(\ref{Wiener-noises}), and keeping terms up to the order $dt$, we obtain
\ba
\langle d\mathcal S_{\rm env}^{\rm ex} \rangle 
&=& \left[ \frac{T}{\gamma} (\partial_i U)^2
- \left(\partial_i  \frac{T}{\gamma}\right) (\partial_i U)
- \frac{T}{\gamma}  (\partial_i^2U) \right] \, dt
\nonumber\\
&=& {-} (\partial_i - (\partial_i U))  \frac{T}{\gamma}  (\partial_i U) \, dt. 
\ea
Multiplying both sides by $p(\xv,t)$ and integrating over $\xv$, we obtain the excess change of environmental entropy at the ensemble level:
\ba
d S_{\rm env}^{\rm ex} = 
\langle  \! \langle d\mathcal S_{\rm env}^{\rm ex} \rangle \! \rangle
=  {-}dt  \int_\xv p (\partial_i - (\partial_i U))  \frac{T}{\gamma}  \partial_i U.
\label{dS-env-0}
\ea 
Integrating by parts once, we transform Eq.~(\ref{dS-env-0}) into:
\ba
 dS_{\rm env}^{\rm ex} 
 = dt \int_{\xv} (\partial_i U) \frac{T}{\gamma} 
 ( \partial_i  +  \partial_i U) \pdf. 
\label{dS-env}
\ea


The system entropy is given by the usual Gibbs-Shannon expression:
\ba
 S_{\rm sys} = - \int_\xv p \log p. 
\ea
Its differential is given by
\ba
d S_{\rm sys} 
&=& - dt \int_{\xv} (\log \pdf) \, \partial_t \pdf
\nonumber\\
&=& - dt  \int_{\xv} (\log \pdf) \partial_i \frac{T}{\gamma} 
\left( \partial_i+ \left(\partial_i U \right) \right) \pdf
\label{dS-sys-0}\\
&=&  dt  \int_{\xv}  (\partial_i \pdf)  \frac{T}{\gamma \,p} 
\left( \partial_i+ \left(\partial_i U \right) \right) \pdf,
\label{dS-sys}
\ea
where in the second equality, we have used Eq.~(\ref{FPE}), whilst in the last equality, we have integrated by parts once.  

The {\em excess EP}~\footnote{This is called the {\em non-adiabatic EP} by Esposito and van der Broeck.  Here we follow the terminology of Glansdorff and Prigogine.} at ensemble level is defined as the sum of Eqs.~(\ref{dS-env}) and (\ref{dS-sys}), which is easily seen to be non-negative:
\ba
d S^{\rm ex} &=& 
dS_{\rm env}^{\rm ex} + dS_{\rm sys}  
\nonumber\\
&=& dt  \int_\xv \frac{T}{ \gamma \pdf}
 \left[ \left( \partial_i+ \left(\partial_i U \right) \right) \pdf \right] ^2 
\geq 0.  \quad
\label{dS-tot}
\ea
This inequality may be called the {\em excess Clausius inequality}. 


We may also define a dimensionless functional 
\ba
\Phi [p] &=& \int_{\xv} p (\xv) [U(\xv) + \log p(\xv)]
\nonumber\\
&=& \int_\xv p(\xv, t) \log \frac{p(\xv, t)}{p_{\rm ss}(\xv)}
= D(p ||p_{\rm ss}),
\label{Phi-def}
\ea
which is the relative entropy between the pdf and the steady state pdf.  Being non-negative and minimized at the steady state, $\Phi [p]$ is in fact the Lyapunov functional of the Fokker-Planck dynamics.  Furthermore, for fixed control parameter, it is easy to see that the rate of $\Phi[p]$ is precisely negative the rate of excess EP:
\ba
- \frac{d\Phi}{dt} = \frac{d S^{\rm ex}  }{dt} \geq 0.
\ea

\subsection{Limit of vanishing temperature gradient}

For the special case of vanishing temperature gradient, both $T$ and $\gamma$ are fixed constants.  We will show that the theory developed above restores to the well known theory of stochastic thermodynamics for Brownian motion in an equilibrium fluid.  For simplicity we assume that the interaction between the Brownian particle and the fluid has no effect on the equilibrium pdf.   

Firstly, in the absence of temperature gradient,  the equilibrium pdf is:
\ba
p_{\rm eq}(\xv) = e^{- \beta (V(\xv) - F)}, 
\ea
where $F = - T \log \int_\xv e^{-\beta V(\xv)}$ is the equilibrium free energy.  Comparing this with Eq.~(\ref{p_ss-U-1}), we see that 
\ba
U(\xv) = \beta (V(\xv) - F). 
\label{U-beta-V-F}
\ea
Using this to replace $U$ in terms of $V$, we find that the Langevin equation (\ref{od-Ito-Langevin}) reduces to the familiar form of over-damped dynamics of a Brownian particle coupled to a single heat bath with temperature $T$:
\ba
 dx^i =- \frac{1}{\gamma } \partial_i V(\xv)dt
 + \sqrt{ \frac{2 T}{\gamma }}dW^i.
\quad \label{od-Ito-Langevin-const-T}
\ea
Now following the procedure in Ref.~\cite{covariant-ST-2021}, we define $V$ as the fluctuating internal energy, and define the differential heat and work at the trajectory level as
\ba
\dbar {\mathscr Q} &\equiv& d_\xv V(\xv; \lambda), \\
\dbar {\mathscr W} &\equiv& d_\lambda V(\xv; \lambda).
\ea
Since the fluid is in equilibrium, the entropy change of the fluid is related to the differential heat via
\ba
d{\mathcal S}_{\rm env} \equiv - \beta \dbar  {\mathscr Q} 
=  - \beta \dbar  V = - d_\xv U, 
\ea
which agrees with Eq.~(\ref{env-ep}).  The inequality (\ref{dS-tot})  then reduces to 
\ba
d S^{\rm tot} = dS_{\rm sys} + dS_{\rm env} 
= dS_{\rm sys} -  \beta \dbar Q \geq 0,
\ea 
which is precisely the usual Clausius inequality.  Finally the functional defined in Eq.~(\ref{Phi-def}) reduces to 
\ba
\Phi [p]  &=& \beta \int_\xv p ( V + T \log p) - \beta F
\nonumber\\
&=& \beta \left( F[p] - F\right),
\ea
where $F[p]$ is the non-equilibrium free energy defined as
\ba
F[p] \equiv \int_\xv p (V + T \, \log p).
\ea
Hence $T \, \Phi$ is the deviation of the non-equilibrium free energy from its equilibrium value. 


\subsection{Fluctuation Theorem for excess EP} 
Let $P(\xv'|\xv; dt)$ be the pdf that the system evolves to state $\xv' = \xv + d \xv$ at time $dt$, given that it is in state $\xv$ at time $0$~\footnote{{We note that both $d\xv$ and $dt$ are infinitesimal quantities.}}.  $P(\xv|\xv'; dt)$ is then the pdf  of the backward process $\xv' \rightarrow \xv$. 
Invoking Eq.~(A27) of  Ref.~\cite{covariant-ST-2021} we obtain~\footnote{note that in the present case both $\xv$ and the control parameter $\lambda$ are even under time-reversal, hence $\xv =\xv^*, \lambda= \lambda^*$.} :
\begin{subequations}
\ba
\log \frac{ P (\xv' | \xv ;dt)}{ P (\xv | \xv';dt)}
= - U(\xv') + U(\xv) = - d_\xv U(\xv). \quad
\label{LDB}
\ea
which, in view of Eq.~(\ref{env-ep}), may be rewritten into
\ba
\log \frac{ P (\xv' | \xv ;dt)}{ P (\xv | \xv';dt)} 
= d\mathcal S_{\rm env}^{\rm ex}. 
\label{LDB-1}
\ea
\label{LDB-1-2}
\end{subequations}
\!\!\!\!\! This establishes the connection between the Langevin dynamics and thermodynamics.  We shall call Eqs.~(\ref{LDB}) and (\ref{LDB-1}) the condition of {\em local detailed balance}. 
Provided that the steady state (\ref{p_ss-U-1}) exists, Eq.~(\ref{LDB}) is in fact equivalent to the condition of detailed balance:
\ba
P (\xv' | \xv ;dt) \, e^{-U(\xv)} 
=  P (\xv | \xv';dt) \, e^{-U(\xv')} . 
\ea

We define the {\em forward process} with a dynamic protocol $\lambda(t), 0 \leq t \leq \tau$, where the system starts from the steady state corresponding to the initial parameter $\lambda(0) = \lambda_i$:
\begin{subequations}
\label{ini-F-B}\label{ini-app}
\ba
p_F(\xv, 0) = e^{- U(\xv,\lambda_i)}. 
\label{ini-F}
\ea
The {\em backward process} is defined by the protocol $\tilde \lambda (t) = \lambda(\tau - t), 0 \leq t \leq \tau$, where the system starts from the steady state corresponding to the initial parameter $\tilde \lambda (0) = \lambda(\tau) = \lambda_f$:  
\ba
p_B(\xv,  0) = e^{- U(\xv,\lambda_f)}.
\label{ini-B}
\ea
\end{subequations}
In both the forward and backward processes, the temperature gradient remains fixed. 

Now consider a {\em forward trajectory} $\xv(t)$, with initial state $\xv(0)$. To simplify notations, we shall use $\bm \gamma$ (bold phase) to denote the forward trajectory, which should be carefully distinguished from the friction coefficient $\gamma$.  The corresponding {\em backward trajectory} is defined as $\tilde \xv (t) \equiv \xv (\tau - t) \equiv \tilde \gammam$. The initial state $\tilde \xv(0) = \xv(\tau)$ of the backward trajectory is identical to the final state of the forward trajectory and vice versa.  We can break both the forward and backward trajectories into a large number of small segments, and apply Eqs.~(\ref{LDB-1-2}) to each pair of segments. Adding up all these results, we obtain
\ba
\log \frac{p_F[\boldsymbol \gamma |  \xv(0)]}
{p_B[\tilde\gammam | \tilde \xv(0)]}
 = - \int_\gammam  d_\xv U
 =  \Delta \mathcal S_{\rm env, F}^{\rm ex}[ {\gammam}].
 \label{log-P-P-2}
\ea
For a careful derivation of this result, see Ref.~\cite{covariant-ST-2021}.  The r.h.s. of the  equation is the excess change of environmental entropy along the entire forward trajectory $\gammam$ in the forward process.  In the l.h.s., $p_F[\boldsymbol \gamma |  \xv(0)]$ and $p_B[\tilde\gammam | \tilde \xv(0)]$ are the pdfs of forward and backward trajectories conditioned on their initial states. 

The backward of the backward process is the forward process, whilst the backward of the backward trajectory is the forward trajectory.  Hence analogous to Eq.~(\ref{log-P-P-2}), we also have
\ba
\log \frac{p_B[\tilde\gammam | \tilde \xv(0)]}
 {p_F[\boldsymbol \gamma |  \xv(0)]}
 =  \int_{\tilde\gammam } d_\xv U
 =  \Delta \mathcal S_{\rm env, B}^{\rm ex}[ {\tilde\gammam}].
 \label{log-P-P-2-1}
\ea
Equation (\ref{log-P-P-2-1}) is the opposite of Eq.~(\ref{log-P-P-2}), hence we find 
\ba
 \Delta \mathcal S_{\rm env, F}^{\rm ex}[ {\gammam}]
 = -  \Delta \mathcal S_{\rm env, B}^{\rm ex}[ {\tilde\gammam}]
  = \int_{\gammam } d_\xv U.
 \label{Sigma-ex-F-def}
\ea

Using the initial state pdfs (\ref{ini-app}), the unconditional pdfs of the forward and backward trajectories can be found:
\begin{subequations}
\label{p_F-p_B-1}
\ba
p_F[\boldsymbol \gamma] &=& 
p_F[\boldsymbol \gamma |  \xv(0)] \,
 e^{- U(\xv(0),\lambda_i)}, \\
p_B[ \tilde{ \boldsymbol \gamma}] &=& 
p_B[\tilde\gammam | \tilde \xv(0)] \,
 e^{- U(\tilde \xv(0),\lambda_f)}.
 \ea 
\end{subequations}
Taking the ratio, and using Eq.~(\ref{log-P-P-2}), we obtain
\ba
\log \frac{p_F[\boldsymbol \gamma]}{p_B[\tilde\gammam ]} 
 =   U(\xv_f,\lambda_f) - U( \xv_i,\lambda_i)
 - \int_\gammam  d_\xv U.
\label{asdf-1}
\ea
Whilst the last term in the r.h.s. is the excess change of environmental entropy along the forward trajectory of the forward process, the first two terms in the r.h.s. may be understood as the change of {\em stochastic entropy}~\cite{Seifert-2005}  of the system along the forward trajectory of the forward process. Combining these two contributions, we define the r.h.s. of Eq.~(\ref{asdf-1}) as {\em the excess EP along the forward trajectory of the forward process}, denoted as $\Sigma^{\rm ex}_{\rm F}[\gammam]$.  It immediately follows that the excess EP along the backward trajectory of the backward process is the negative of Eq.~(\ref{asdf-1}), i.e. $\Sigma^{\rm ex}_{\rm B}[\tilde \gammam] = - \Sigma^{\rm ex}_{\rm F}[\gammam]$. 

Since the difference $U(\xv_f,\lambda_f) - U( \xv_i,\lambda_i)$ can be written as the integral $\int_\gammam   d U$, which can be further written as $\int_\gammam  (d_\xv U + d_\lambda U)$ up on invoking Eq.~(\ref{dU-expand}), we may also write Eq.~(\ref{asdf-1}) as:
\ba
 \Sigma^{\rm ex}_{\rm F}[\gammam]
 = - \Sigma^{\rm ex}_{\rm B}[\tilde \gammam]
 = \int_\gammam  d_\lambda U 
 =  \log \frac{p_F[\boldsymbol \gamma]}
{p_B[\tilde\gammam ]} .
 \label{Sigma-ex-F}
\ea 
which may be called  {\em the detailed FT for the excess EP}.

The pdfs of {excess EP} of the forward and backward processes are respectively:
\begin{subequations}
\ba
&&p_F( \Sigma^{\rm ex}) \equiv \int D\gammam  \, p_F(\gammam ) \, \delta 
\left(  \Sigma^{\rm ex} - \log \frac{p_F[\gammam ]}{p_B[\tilde \gammam ]} \right),
\quad \quad \\
&&p_B( \Sigma^{\rm ex}) \equiv \int D\tilde \gammam  \, p_B(\tilde \gammam )
\,  \delta \left(  \Sigma^{\rm ex} + \log \frac{p_F[\gammam ]}{p_B[\tilde \gammam ]} \right).
\ea
\end{subequations}
The functional integrals over trajectories should be calculated using time-slicing, much similar to the Feynman path integral in quantum mechanics.  Using Eq.~(\ref{Sigma-ex-F}), we can then show 
\ba
p_F( \Sigma^{\rm ex}) &=& 
\int \!\! D\gammam  \, p_F(\gammam ) \, \delta 
\left(  \Sigma^{\rm ex} - \log \frac{p_F[\gammam ]}{p_B[\tilde \gammam ]} \right)
\nonumber \\
& = & \int  \!\! D \gammam  \, p_B( \tilde \gammam ) \, e^{ \Sigma^{\rm ex} } \, \delta 
\left(  \Sigma^{\rm ex} - \log \frac{p_F[\gammam ]}{p_B[\tilde \gammam ]} \right) \nonumber\\
& = & e^{ \Sigma^{\rm ex} } \int D \tilde \gammam  \, p_B(\tilde \gammam ) \delta 
\left(  \Sigma^{\rm ex}
 + \log \frac {p_B[\tilde \gammam ]} {p_F[\gammam ]} \right) \nonumber\\
& = & e^{ \Sigma^{\rm ex} }  p_B(-  \Sigma^{\rm ex}).
\ea
Hence we arrive at 
\ba
\log \frac{p_F( \Sigma^{\rm ex})}{p_B(- \Sigma^{\rm ex})} 
= { \Sigma^{\rm ex}}. 
\label{FT}
\ea
which is called {\em the transient FT for the excess EP}. 
 
In the limit of vanishing temperature gradient, we may use Eq.~(\ref{U-beta-V-F}) and Eq.~(\ref{Sigma-ex-F}) to obtain 
\ba
\Sigma_F [\gammam] &=& \beta \int_\gammam d_\lambda  \left( V(\xv; \lambda) - F(\lambda) \right)
\nonumber\\
&=& \beta {\mathcal W}[\gammam]
- \beta \Delta F,
\ea
where $\Delta  F = F (\lambda_f) - F(\lambda_i)$ is the change of the equilibrium free energy, whilst  ${\mathcal W}[ {\gammam}] = \int_{\gammam} d_\lambda  V(\xv; \lambda)$ is the work done by the external parameter. Hence the excess FT (\ref{FT}) reduces to the well-known Crooks FT:
\ba
\log \frac{p_F(W)}{p_B(- W)} = \beta (W - \Delta F). 
\ea

\section{ Numerical simulation} 
\label{sec:simulation}
In this section, we simulate a Brownian particle in a temperature gradient, and verify that the dynamics of the Brownian particle is indeed described by the over-damped Langevin equation (\ref{od-Ito-Langevin}), with appropriate choices of system parameters.  We will numerically compute the position-dependent temperature $T(\xv)$ and friction coefficient $\gamma(\xv)$, as well as the generalized potential $U(\xv; \lambda)$.   Using these results, we explicitly verify the FT (\ref{FT}), which turns out to hold not only in the over-damped regime, but also in the under-damped regime.

\vspace{-3mm}
\subsection{Setup of simulation}

We simulate a Brownian particle immersed in a 2D  Lennard-Jones  fluid using LAMMPS package~\cite{plimpton1995fast} of Molecular Dynamics (MD).  

Both the interaction between fluid particles, as well as that between the Brownian particle and fluid particles, are modeled by the 12-6 Lennard-Jones (LJ) pair potential truncated at zero potential.  Throughout this section all lengths are in units of the Lennard-Jones diameter $\sigma_0$, all energies are in units of the well-depth $\epsilon_0$, and the unit of time is $\tau=\sqrt{m_0\sigma_0^2/\epsilon_0}$, where $m_0$ is the mass of the fluid particles. Boltzmann's constant is set to unity.   The interaction potential between two fluid particles is
\begin{equation}
V_{\rm ff}(r)=\left\{
    \begin{aligned} &
         4  \left[  
      \left( \frac{ 1 }{r} \right)^{12}
      -    \left(  \frac{1 }{r} \right)^6 \right], 
      & r< 1, \\
    &    0, &  r \geq 1. \\
    \end{aligned}
    \right.
\end{equation}
 The mass of the Brownian particle is set to be $M = 10 m_0$.   The diameter of the Brownian particle is chosen to be triple that of the fluid particles. Consequently, the interaction potential between the Brownian particle and a fluid particle is
\begin{equation}
V_{\rm cf}(r)=\left\{
    \begin{aligned} &
         4  \left[  
      \left( \frac{ 2 }{r} \right)^{12}
      -    \left(  \frac{ 2  }{r} \right)^6 \right], 
      & r< 2,  \\
    &    0, & r \geq 2. \\
    \end{aligned}
    \right.
\end{equation}

\begin{figure}[t!]
    \centering
    \includegraphics[width= 3.3in]{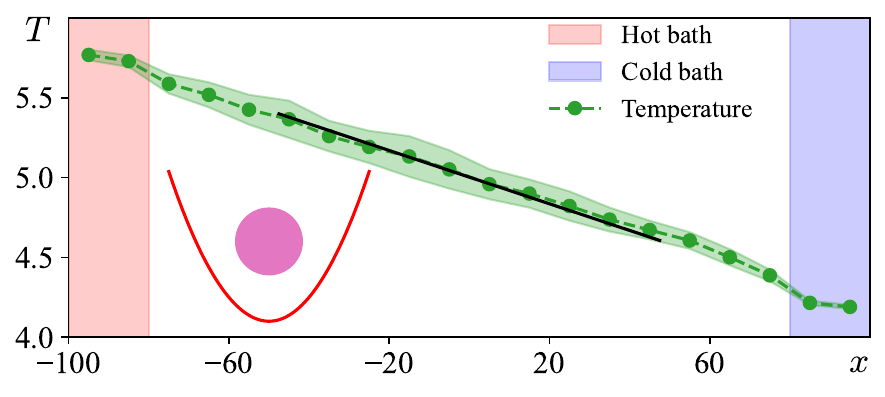}
    \caption{Brownian particle in temperature gradient and external potential.  The green dots show the fluid temperature determined numerically, whereas the green shade gives the error bar. The straight-line is the linear fit of temperature given by Eq.~(\ref{T-linear-fit}). }
    \label{fig::grad1/temp_xpdf}
 \vspace{-5mm}
\end{figure}

The system consists of $N=2810$ fluid particles and a single Brownian particle.  All particles are put into a two-dimensional square box with length $L=200$. The dimensionless density of the fluid is $\pdf_{e}=0.07$, which corresponds to a gas phase.    The NVE ensemble is used to carry out the MD simulation.  A temperature gradient is established using the Langevin thermostat method~\cite{schneider1978molecular}, by introducing a hot region with temperature $T_h = 5.8$ near the left boundary, and a cold region with temperature $T_c = 4.2$ near the right boundary. We choose the reflection boundary condition at $x=\pm100$, so that there is no macroscopic transport of fluid particles in the steady state.  We choose the periodic boundary condition at $y = \pm 100$. See Fig.~\ref{fig::grad1/temp_xpdf} for an illustration.

We choose the time step of MD integration as $dt=0.001$, and start the simulation from a random state.  After running $6\times 10^5$ steps, a steady temperature gradient is set up. 



\begin{figure}[t!]
    \centering
    \includegraphics[width=3.4in]{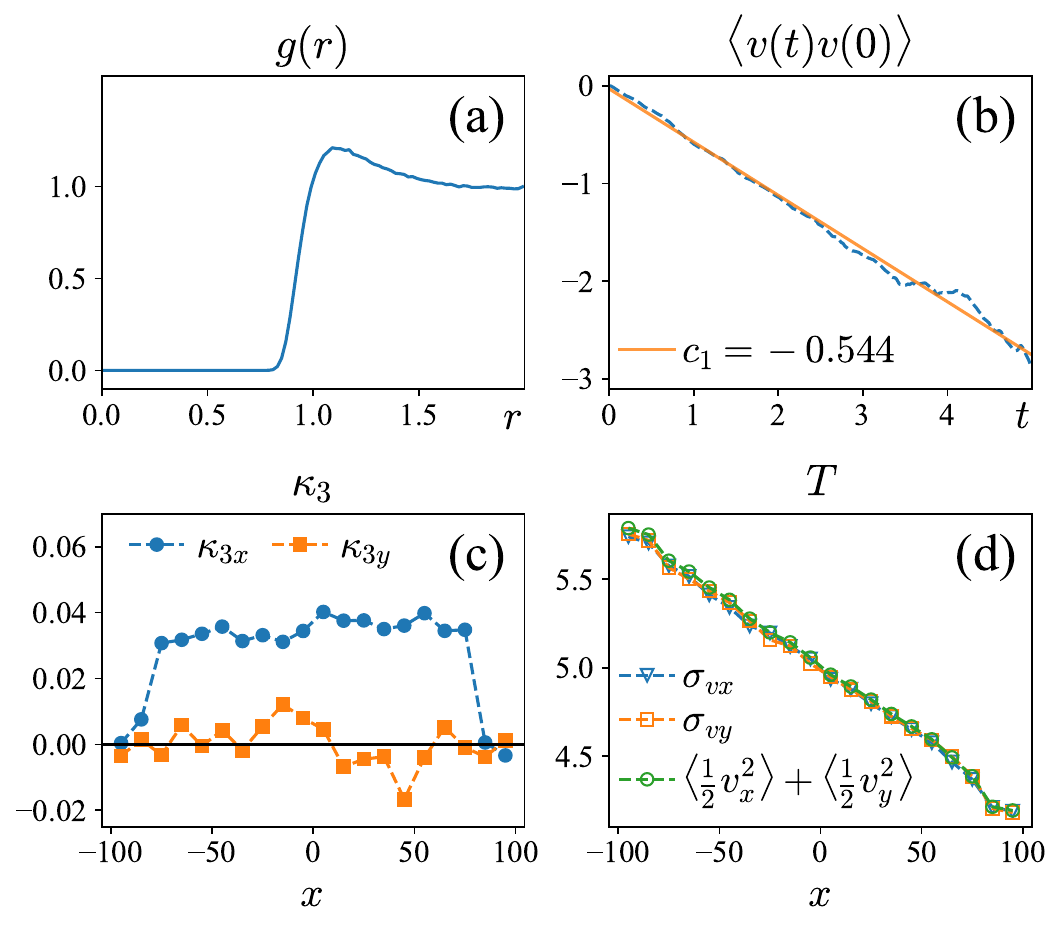}
    \vspace{-5mm}
    \caption{(a) and (b) refer to equilibrium fluid. (a)  Radial distribution function (RDF)  of fluid particles; (b) log-linear pro of normalized velocity auto-correlation function (VACF) of fluid particles, where \wu{$c_1$ is the slope of the linear fit}.  (c) and (d) refer to fluid with temperature gradient. (c) Third order cumulants of $v_x, v_y$ as functions of $x$; (d) temperature as a function of $x$ computed use three different methods.  The meaning of vertical axis is shown on the top of each panel. }
    \label{fig::combined/env_rdf_vacf_k3m_T}
 \vspace{-3mm}
\end{figure}

To calibrate the simulation, we first simulate an equilibrium fluid by choosing $T_h = T_c = 5$, without inserting the Brownian particle. The radial distribution function (RDF) of fluid particles as a function of $r$ is shown in Fig.~\ref{fig::combined/env_rdf_vacf_k3m_T} (a).  The normalized velocity auto-correlation function (VACF) as a function of $t$ is shown in Fig.~\ref{fig::combined/env_rdf_vacf_k3m_T} (b) in log-linear scale.  It can be fit well by an exponential:
\begin{equation}
\frac{\langle v(t)v(0)\rangle}{\langle v(0)^2  \rangle } 
 =  e^{-\gamma_0 t/m_0}.
\end{equation}
These results  indicate the equilibrium nature of the fluid.

We then introduce the temperature gradient.  We cut the system into many slices along the $x$ axis, with thickness $\Delta x = 5, {\rm or} \,10$, both of which are longer than the mean free path of fluid particles.   Within each slice, the velocity distribution of fluid particles is approximately Gaussian. Deviations from Gaussian may be characterized by the normalized third-order cumulants:
\ba
    \kappa_{3x} &=& \frac{\langle (v_x- \langle v_x \rangle)^3 \rangle}
    {\langle (v_x- \langle v_x \rangle)^2 \rangle^{3/2}}, 
    \label{kappa_3x-def}\\
    \quad \kappa_{3y} &=& \frac{\langle (v_y - \langle v_y \rangle)^3 \rangle}
    {\langle (v_y - \langle v_y \rangle)^2 \rangle^{3/2}}.
\ea
These cumulants are computed in each slice and shown in Fig.~\ref{fig::combined/env_rdf_vacf_k3m_T} (c).  As one can see there, $\kappa_{3y}$ fluctuates around zero, whereas $\kappa_{3x}$ exhibits systematic deviation from zero by a few percent.  The latter is evidently related to the heat transport along $x$ direction in the steady state.   Hence the pdf of $v_y$ is Gaussian, whereas the pdf of $v_x$ exhibits a small yet systematic deviation from Gaussian.

We compute the temperature of each slice using three different methods:
(1) average kinetic energy $T=\langle \frac{1}{2} v_x^2 \rangle + \langle \frac{1}{2} v_y^2 \rangle $, (2) variance of $v_x$, and (3) variance of $v_y$.  As shown in Fig.\ref{fig::combined/env_rdf_vacf_k3m_T} (d), all three methods are consistent with a constant temperature temperature.  The best fit of the temperature profile is  
\ba
T(x)=-0.0083 \, x+5.005.  
\label{T-linear-fit} 
\ea


\subsection{Velocity distribution of the Brownian particle}
\label{sec:property-brownian-particle}

\begin{figure}[t!]
    \centering
    \includegraphics[width=3.4in]{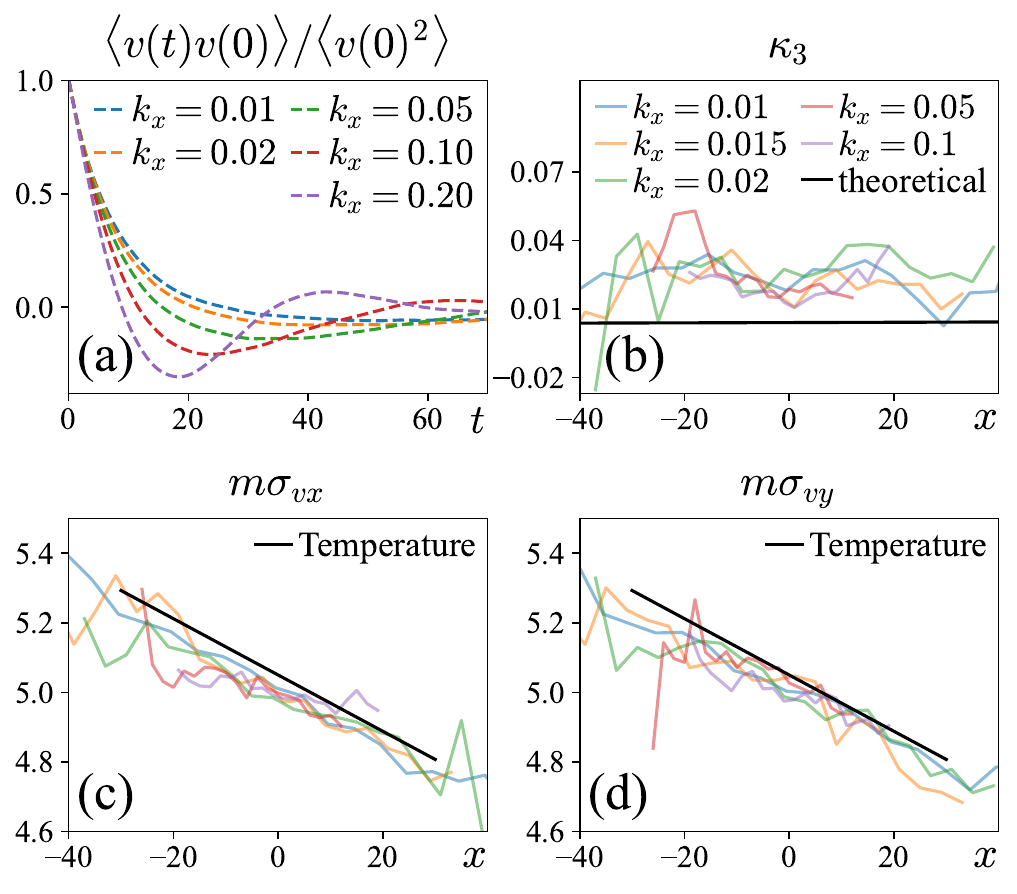}
    \vspace{-5mm}
    \caption{(a) The VACF of the Brownian particle in $x$ direction at difference external potential $k_x$; (b): third order cumulant of $v_x$ of the Brownian particle as a function of $x$.  The black straight-line is {computed using the Eq.~(15) of Ref.~\cite{Celani2012}. }  
    (c) and (d): Variances of $v_x, v_y$ can be used to compute the local temperature. The parameters corresponding to the colored lines are the same as in (b). The black straight-line is the linear fit Eq.~(\ref{T-linear-fit}).
    The  meaning of vertical axis is shown on the top of each panel. 
    }
        \label{fig::combined/kappa_v_U_fit}
\end{figure}

A Brownian particle is inserted, which also couples to an anisotropic harmonic potential:
\ba
V(\xv)= \frac{1}{2}k_x  (x-x_0)^2 
+  \frac{1}{2}k_y  y ^2. 
\label{equ::potential}
\ea
First we set $k_y = x_0 = 0$, and compute the VACF of Brownian particle in $x$ direction, for different values of $k_x$.  As shown in Fig.~\ref{fig::combined/kappa_v_U_fit} (a), for $k_x > 0.05$, the VACF decays in an oscillatory fashion, which means that the system is in the under-damped regime. By contrast for $k_x < 0.05$, the VACF decays without oscillation, which means that the system is in the over-damped regime.  Hence we expect that for $k_x < 0.05$, the Brownian dynamics is adequately described by the over-damped Langevin equation (\ref{od-Ito-Langevin}), whereas for $k_x > 0.05$, the dynamics should be described by under-damped Langevin equation. 

Next we compute the velocity pdf of the Brownian particle in each $x$-slice.  Similar to the fluid particles, the normalized third order cumulant $\kappa_{3x} $ of the Brownian particle also deviate slightly but systematically from zero, as shown in Fig.~\ref{fig::combined/kappa_v_U_fit}(b).  Furthermore, $\kappa_{3x}$ seems independent of confining potential and of the slice, and is also systematically different from the prediction of Ref.~\cite{Celani2012}.

We also plot the variances of velocity components $v_x, v_y$ of the Brownian particle as a function of $x$.  As shown in Fig.~\ref{fig::combined/kappa_v_U_fit}(c) and (d), the results are consistent with the temperature profile Eq.~(\ref{T-linear-fit}), which is computed using the velocity distributions of the fluid particles.  This verifies that the Brownian particle establishes local thermal equilibrium with the fluid within each slice.

\subsection{Position distribution of the Brownian particle}
\label{sec:property-brownian-particle-U}

\begin{figure}[t!]
    \centering
    \includegraphics[width=3.4in]{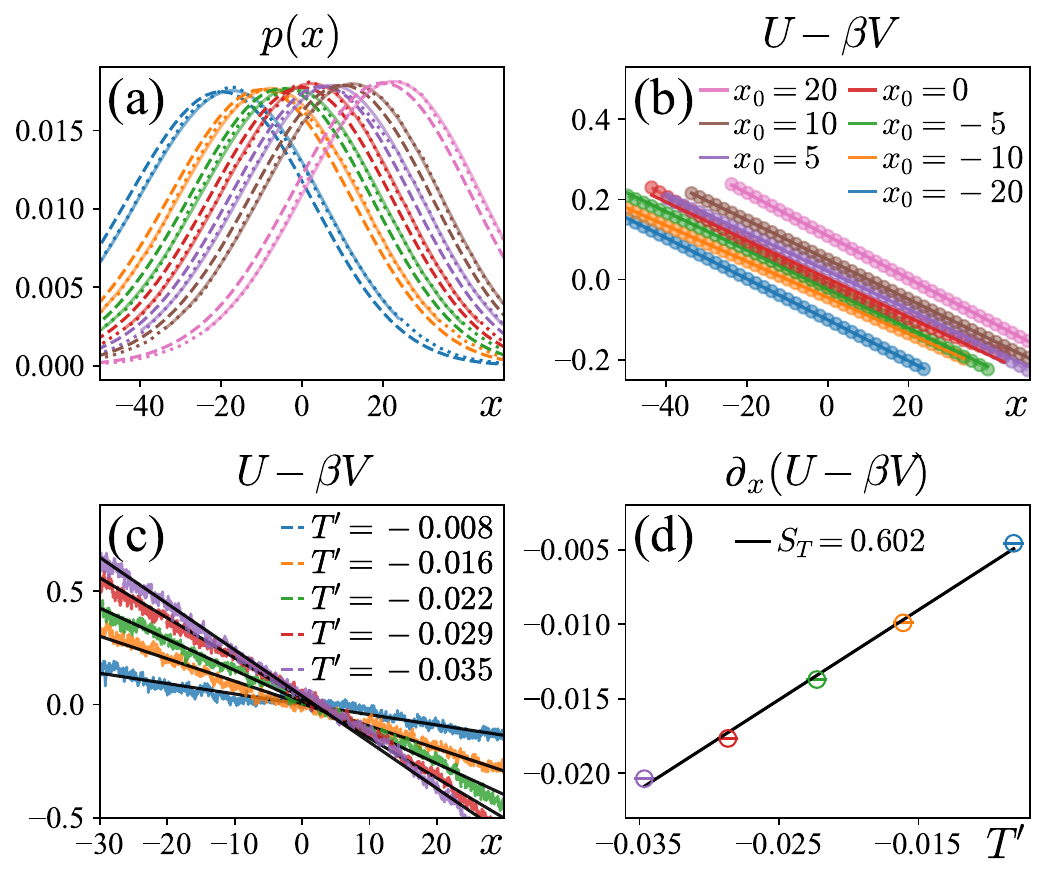}
    \vspace{-5mm}
    \caption{
    (a) Numerically measured steady state pdf  (dotted line), $\exp(-U_{\rm fit})$ (solid line), and $\exp(-\beta V)$ (dashed line); (b) $(U_{fit}-\beta V)$; (c) $U - \beta V$ against $x$ for different values of temperature gradient. The black straight-lines are linear fitting.  (d): The slope of the linear fit against the temperature gradient gives the Soret coefficient. $k_x =0.01$ in all these simulations. The  meaning of vertical axis is shown on the top of each panel. }
        \label{fig::combined/U_fit_gd}
  \vspace{-3mm}
\end{figure}

We compute the steady state position pdf of Brownian particle,  and use it to obtain the generalized potential via Eq.~(\ref{p_ss-U-1}).  The results are shown as dotted lines in Fig.~\ref{fig::combined/U_fit_gd}(a) for different values of $x_0$.  Also shown there are $e^{-\beta (x) V(x)} $ (dashed lines, normalized properly), which appear always to the left of the steady state pdf.  This indicates that the thermophoresis drives Brownian particle to the right (with lower temperature).  


We fit the steady state position distribution by $e^{-  U_{\rm fit}(\xv; \lambda)} $, where $U_{\rm fit}(\xv; \lambda) $ is given by
\ba
U_{\rm fit}(\xv; \lambda) =
 \beta(\xv) V(\xv; \lambda) + S_T\, \xv \cdot \nabla T 
+ b(\lambda),  
\label{U-V-S_T}
\ea
with $T' = - 0.0083$ as given by Eq.~(\ref{T-linear-fit}), $S_T$ is the  fitting parameter, whilst $b$ is a normalization constant.  As shown by solid lines in Fig.~\ref{fig::combined/U_fit_gd}(a), the fitting quality is remarkably good.  


 In Fig.~\ref{fig::combined/U_fit_gd}(b) we plot $U_{\rm fit} - \beta V$ as a function of $x$, for fixed $k_x$  but different values of $x_0$.  The slope of these straight-lines, which can be understood as the effective thermophoresis force, renormalized by the local temperature $T(x)$, are approximately independent of $x_0$.  This again verifies the validity of Eq.~(\ref{U-V-S_T}). 
  
\subsection{Soret coefficient}

Let us now establish the connection between our theory and the thermodynamic theory of thermophoresis~\cite{Piazza-review-2008}.  In the latter theory, there is no external potential, the probability current of the Brownian particle is~\cite{Piazza-review-2008}
\be
{\boldsymbol j} (\xv) = - D \left( \nabla p(\xv) + S_T\, p(\xv)   \, \nabla T \right),
\label{j_1}
\ee
where $S_T$ is the {\em Soret coefficient}, whilst $D$ is the diffusion constant.  In general, both $S_T$ and $D$ depend on position.  However, in the thermodynamic theory of thermophoresis, it is always assumed that the overall variation temperature is small, so that $S_T$ and $D$ may be treated as constants.

In our theory, the probability current is given by Eq.~(\ref{j_2}), where $U$ is given by Eq.~(\ref{U-V-S_T}).  In the absence of external potential, Eq.~(\ref{j_2}) reduces to
\ba
{\boldsymbol j} (\xv) = - \frac{T(\xv)}{\gamma(\xv)} 
\left( \nabla p(\xv) + S_T \, p(\xv) \, \nabla T \right).
\ea
This is precisely Eq.~(\ref{j_1}) with a position-dependent diffusion constant $D(\xv) = T(\xv)/\gamma(\xv) $.  Hence the parameter $S_T$ in Eq.~(\ref{U-V-S_T}) is precisely the {\em Soret coefficient}.  

In Fig.~\ref{fig::combined/U_fit_gd} (c) we fix $x_0= 0$, and use simulation data to plot $U - \beta V$ against $x$ for different values of  temperature gradient.  In Fig.~\ref{fig::combined/U_fit_gd} (d)  we plot the slope of linear fitting in Fig.~\ref{fig::combined/U_fit_gd} (c) against temperature gradient.  This clearly demonstrates that the Soret coefficient $S_T$ is approximately constant within the region of simulation.  Indeed, in the region of simulation, the overall variation of temperature is less than 8\%, hence $S_T$ can be treated approximately as a constant.




\subsection{Friction coefficient}
\begin{figure}[t!]
    \centering
    \includegraphics[width=3.3in]{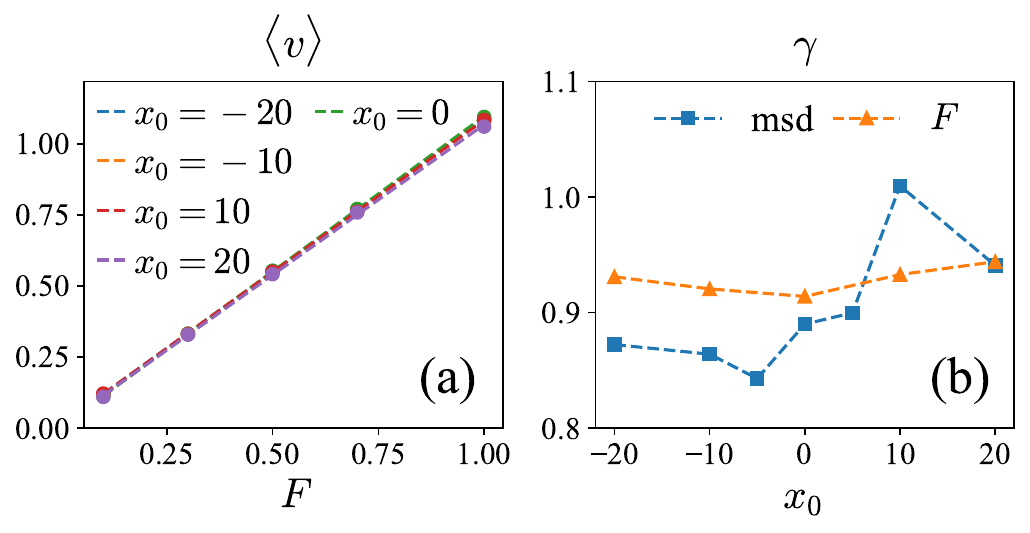}
  \vspace{-5mm}
     \caption{
(a) The linear relation between force $F$ and average velocity is used to calculate the friction coefficient $\gamma(x)$. It is seen that $\gamma(x)$ is only very weakly dependent on  $x$.  (b) The friction coefficient $\gamma$ computed using two different methods. $k_x=0.01$ in all these simulations. }
    \label{fig::combined/V0_free_vacf_k_gamma}
\end{figure}





We use two different methods to compute the friction coefficient as a function of $x$. 
In the first method, we apply a constant force along $y$ direction, and compute the  friction coefficient via:
\begin{equation}
    {\gamma(x_0)} \langle v_y(t)\rangle = {F},
\end{equation}
where $x_0$ is the minimum of the potential Eq.~(\ref{equ::potential}). In Fig.~\ref{fig::combined/V0_free_vacf_k_gamma} (a) the relation between $ \langle v_y(t)\rangle$ and  $F$ is shown for different $x_0$, from which we deduce that the dependence of ${\gamma(x_0)} $ on $x_0$ is very weak.   The computed friction coefficient is shown as orange symbols in  Fig.~\ref{fig::combined/V0_free_vacf_k_gamma} (b).

In the second method, we set $k_y = 0$, so that the Brownian particle diffuses freely along $y$ direction, whilst it is confined along $x$ direction.  Provided that the Langevin equation (\ref{od-Ito-Langevin}) is applicable, the mean square displacement (MSD) is given by
   \begin{equation}
        \langle \Delta y(t)^2 \rangle = \frac{ 2 T(x_0)}{\gamma(x_0)} t,
        \label{Delta-y^2-1}
    \end{equation}
where $T(x)$ is given by the linear fit Eq.~(\ref{T-linear-fit}).   Using Eq.~(\ref{Delta-y^2-1}) to fit simulation data, we obtain the friction coefficient shown as blue symbols in  Fig.~\ref{fig::combined/V0_free_vacf_k_gamma} (b).  

It can be seen from Fig.~\ref{fig::combined/V0_free_vacf_k_gamma} (b) that whilst two methods are yield generally consistent results, the result of the second method exhibit much larger fluctuations.  According to the first method, the friction coefficient is approximately independent of the position along $x$ direction.  This is rather expected, since variation of temperature in the simulation region is only a few percent.

\subsection{ Verification of FT}
\label{sec:verification-FT}

\begin{figure}[t!]
    \centering
    \includegraphics[width=3.4in]{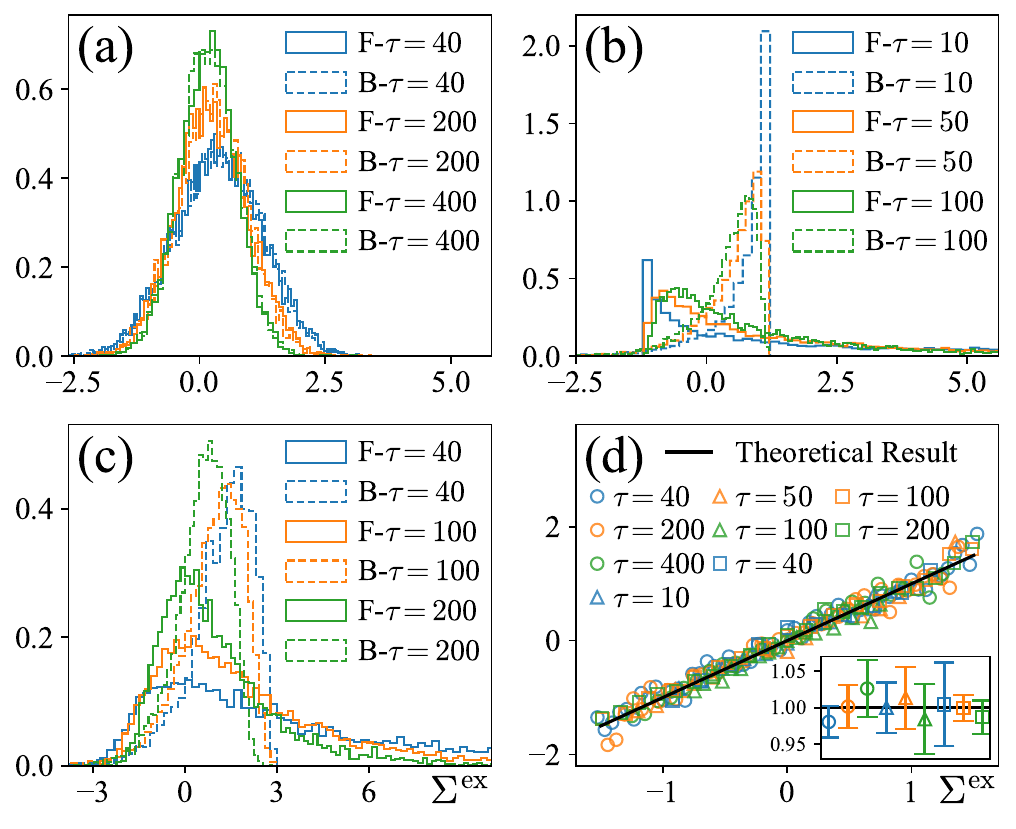}
    \vspace{-5mm}
    \caption{ Verification of excess FT. (a), (b), (c): Histograms of excess EP. In all legends $F,B$ means forward and backward respectively.  The protocol of  (a) is  $x_0=-10+20 t/\tau, k_x =0.01, k_y =0.01, t \in (0, \tau)$. The value of $\tau$ (duration of the process) is shown in the legend.  The protocol of  (b) is  $x_0=0, k_x =0.01+0.1 t/\tau, k_y =0.1$.  The protocol of (c) is  $x_0=-10+20 t/\tau, k_x =0.01+0.1 t/\tau, k_y =0.1$. (d): Verification of FT for all three processes, where circles, triangles, and squares are respectively data from the first, second, and third protocols.    Inset: The fitting slope and estimated error of the FT for each process.  Symbols and colors are as the same as shown in the legend.
    }
    \label{fig::grad1/w_u_grid}
  \vspace{-3mm}
\end{figure}

We simulate three types of forward processes as well as the corresponding backward processes.  In all simulations we fix $k_y = 0.1$, so that the Brownian particle is localized in $y$ direction. In the first protocol, we fix $k_x$ and vary the equilibrium position $x_0$ during the process. In the second protocol, we fix $x_0$ and vary $k_x$.  In the third protocol we vary $k_x, x_0$ simultaneously.  $x_0$, or $k_x$, or both, therefore, are the control parameter $\lambda$ we discussed in Sec. II.  The details of these protocols are explained in the captions of Fig.~\ref{fig::grad1/w_u_grid}. 

For each process, we sample a large number of trajectories.  For each trajectory, we calculate the excess EP using Eq.~(\ref{Sigma-ex-F}).  More specifically, we discretize each trajectory and approximate the excess EP by:
\begin{equation}
    \Sigma^{\rm{ex}}[\boldsymbol \gamma] = \sum_{i} U_{\rm{fit}}(\lambda_{i+1},x_{i})- U_{\rm{fit}}(\lambda_{i},x_{i}),
\end{equation}
where $U_{\rm{fit}}$ is given by Eq.~(\ref{U-V-S_T}).  The pdfs of $\Sigma^{\rm ex}$ in the forward process are then obtained using histograms.  The pdfs of $\Sigma^{\rm ex}$ in the backward process are obtained using similar methods.  The results are shown in Fig.~\ref{fig::grad1/w_u_grid}(a), (b), (c), respectively, for three processes.  In Fig.~\ref{fig::grad1/w_u_grid}(d),  the log ratio $\log p_F(\Sigma^{\rm ex}) /p_B(- \Sigma^{\rm ex})$ is plotted against $\Sigma^{\rm ex}$ for all three processes.   As shown there, all data collapse to the straight-line with unit slope, as predicted by the excess FT Eq.~(\ref{FT}).



\begin{figure}[t!]
    \centering
    \includegraphics[width=3.4in]{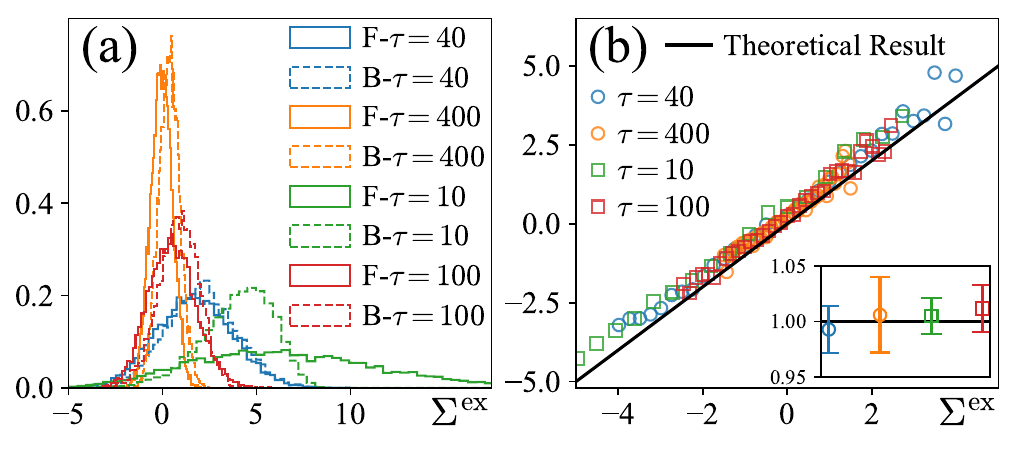}
    \vspace{-7mm}
    \caption{ {Verification of excess FT in the under-damped regime. The protocols of the blue and orange lines are $x_0=-10+20 t/\tau, k_x =0.1, k_y =0.1$, with values of $\tau$ shown in the legend.  The data are shown as circles in (b).  The protocols of the green and read lines are $x_0=-10+20 t/\tau, k_x =0.1+0.1 t/\tau, k_y =0.1$. The data are shown as squares in (b).
    Inset: The slopes of linear fitting and the estimated errors, where symbols and colors are the same as shown in the legend.
    }}
    \label{fig::grad1/w_u_grid_k01}
    \vspace{-3mm}
\end{figure}

As explained in the captions of Fig.~\ref{fig::grad1/w_u_grid} and Fig.~\ref{fig::grad1/w_u_grid_k01}, in the second and third protocols, the spring constant $k_x$ is tuned from $0.01$ to $0.11$. According to the results demonstrated in Fig.~\ref{fig::combined/kappa_v_U_fit} (c), for $k_x \geq 0.05$,  the Brownian dynamics is under-damped, and hence the over-damped Langevin equation (\ref{od-Ito-Langevin}) is not applicable.  Nonetheless, the excess FT seems hold to high precision in these two processes, as can be seen from Fig.~\ref{fig::grad1/w_u_grid}(d).   It then seems that the excess FT Eq.~(\ref{FT}) is valid even in the under-damped regime.  To test this hypothesis,  we also simulate two other protocols where the system is deep in the under-damped regime in the entire process.  As shown in Fig.~\ref{fig::grad1/w_u_grid_k01}, indeed the excess FT  (\ref{FT}) remains valid as well.  



The validity of the excess FT in the under-damped regime may be understood using the stochastic thermodynamic theory developed in Ref.~\cite{nc-2022}, together with two reasonable assumptions.  The Brownian dynamics can be described by certain under-damped Langevin equation.  Even though we do not know the concrete form of this equation, we do expect that it obeys local detailed balance.  The  stochastic thermodynamic theory developed in Ref.~\cite{nc-2022} is then applicable.  (Unlike the over-damped theory as studied in the present work, in the under-damped theory, the house-keeping EP is generically non-vanishing.  This point is however irrelevant, since we are only interested in the excess EP. ) 
The excess EP in the underdamped theory is defined as 
\ba
\Sigma^{\rm ex}_{\rm ud}[\bm \gamma] =
 \int_{\boldsymbol \gamma} d_\lambda U_{\rm ud}(\xv, \pv; \lambda), 
 \label{Sigma-ex-ud-def}
\ea
where $\pv$ is the momentum of the Brownian particle and $U_{\rm ud}(\xv, \pv; \lambda)$ is the generalized potential of the under-damped theory. Here the subscript ud denotes ``under-damped''.  In fact $\Sigma^{\rm ex}_{\rm ud}[\bm \gamma] $ is precisely the functional defined in Eq.~(4.62) of Ref.~\cite{nc-2022}), which is also shown there to obey the fluctuation theorem (c.f. Eq.~(4.63) of Ref.~\cite{nc-2022}):
\ba
\log \frac{p_F( \Sigma^{\rm ex}_{\rm ud})}{p_B(- \Sigma^{\rm ex}_{\rm ud})} 
= { \Sigma^{\rm ex}_{\rm ud}}. 
\label{FT-ed}
\ea

We can decompose the generalized potential $U_{\rm ud}(\xv, \pv; \lambda)$ into the following form:
\ba
U_{\rm ud}(\xv, \pv; \lambda) = U_\Xv(\xv; \lambda) 
+ U_\Pv(\pv; \xv, \lambda), 
\label{U-U_x-U_p}
\ea
such that the conditional steady-state distribution of $\pv$ given $\xv$ and $\lambda$ is given by
\ba
p^{\rm SS}(\pv | \xv, \lambda) 
= \frac{e^{- U(\xv, \pv; \lambda)}}{ \int e^{- U(\xv, \pv; \lambda)} d\pv}
= e^{- U_\Pv(\pv;  \xv, \lambda)}.
\ea
It then follows that the marginal steady-state pdf of $\xv$ is related to $U_\Xv$ via 
\ba
p^{\rm SS}_\Xv(\xv) = e^{- U_\Xv(\xv; \lambda) }.
\ea
Hence $ U_\Xv(\xv; \lambda) $ is precisely the generalized potential appearing in our over-damped theory.  

Now, all results shown in Fig.~\ref{fig::combined/kappa_v_U_fit} indicate that the velocity pdfs of the Brownian particle, conditioned on position variable,  is independent of the control parameters $\lambda$, i.e. $d_\lambda U_\Pv(\pv; \xv, \lambda) = 0$.  From Eq.~(\ref{U-U_x-U_p}) we then deduce 
\ba
d_\lambda U_{\rm ud}(\xv, \pv; \lambda) = d_\lambda U_\Xv(\xv; \lambda), 
\ea
Substituting this back into Eq.~(\ref{Sigma-ex-ud-def}) we find: 
\ba
\Sigma^{\rm ex}_{\rm ud}[\bm \gamma] =
 \int_{\boldsymbol \gamma} d_\lambda U_{\rm ud}
 =  \int_{\boldsymbol \gamma} d_\lambda U_{\Xv}
= \Sigma^{\rm ex}_{\rm od}[\bm \gamma].
 \label{Sigma-ex-ud-def-1-1}
\ea
In other words, the excess EP in the under-damped theory is identical to that in the over-damped theory, provided that (i) the under-damped Langevin theory obeys local detailed balance, and (ii) the velocity distribution of the Brownian particle is independent of the control parameter $\lambda$.  This explains why the excess FT (\ref{FT}) also holds in the under-damped regime. 


\vspace{-3mm}

\section{Conclusion and discussion}
\label{sec:conclusion}
\vspace{-2mm}

In this work, we developed a full theory of stochastic thermodynamics for Brownian motion inside a temperature gradient, and supply systematic numerical verification of the theory. The present work consists a first concrete example of the general theoretical framework developed in {a sequel} of recent papers~\cite{covariant-Langevin-2020,covariant-ST-2021,strong-coupling-2021,nc-2022}.  The problem of Brownian motion in temperature gradient has two distinct features: (i) the system is embedded in a dissipative background, and (ii) the system is coupled to multiplicative noises.  Whilst there have been numerous previous works on this problem, to the best of our knowledge, our work is the first to supply a relatively complete and self-consistent theory of stochastic thermodynamics for this system. 

The over-damped theory of Brownian motion in temperature is simple, because it obeys the conditions of detailed balance, so that the house-keeping entropy production vanishes identically.  Also because both the variables and the control parameters are even under time-reversal, there is no entropy pumping~\cite{nc-2022}.  In the future works, we shall apply the framework developed in Ref.~\cite{covariant-Langevin-2020,covariant-ST-2021,strong-coupling-2021,nc-2022} to more complicated systems, such as Brownian dynamics of rigid bodies, of micro-magnetism, and of Brownian particles in gradient flow.  We are also interested in stochastic thermodynamics of active matters, which have attracted some recent interests~\cite{Fodor2016, Mandal2017}.

The authors acknowledge support from Chenxing Post Doctoral Incentive Project, NSFC grant \#12375035, and Shanghai Municipal Science and Technology Major Project via grant 2019SHZDZX01.

\appendix

\end{document}